\documentclass[journal]{IEEEtran}

\usepackage[bookmarks=true,colorlinks=true,       
    linkcolor=black,          
    citecolor=black,        
    filecolor=black,      
    urlcolor=black           
]{hyperref}

\usepackage{amsfonts}
\usepackage{amsmath}
\usepackage{multirow}

\usepackage{amsfonts}
\usepackage{amsmath}
\usepackage{multirow}

\usepackage{caption}
\usepackage{subcaption}

\usepackage{mathptmx} 
\usepackage{times} 
\usepackage{amsmath} 
\usepackage{amssymb}  

\usepackage[ruled,vlined]{algorithm2e}
\usepackage{algorithmicx}
\usepackage{algpseudocode}
\usepackage{amsthm}

\newtheorem{Problem}{Problem}

\usepackage{cite}

\ifCLASSINFOpdf
   \usepackage[pdftex]{graphicx}
\else
   \usepackage[dvips]{graphicx}
\fi
\usepackage{todonotes}

\begin{document}
%

\title{On the Effects of Distributed Electric Vehicle Network Utility Maximization in Low Voltage Feeders}

%

\author{Jose~Rivera
        ~and~Hans-Arno~Jacobsen
\thanks{Jose Rivera, and Hans-Arno Jacobsen are with Technical University Munich, D-85748, Munich, Germany. (\texttt{j.rivera@tum.de, jacobsen@in.tum.de})}
}
\maketitle

\begin{abstract}
The fast charging of Electric Vehicles (EVs) in distribution networks requires real-time EV charging control to avoid the overloading of grid components. Recent studies have proposed congestion control protocols, which result from distributed optimization solutions of the Network Utility Maximization (NUM) problem. While the NUM formulation allows the definition of distributed computations with closed form solutions, its simple model does not account for many of the feeders operational constraints. This puts the resulting control algorithms effectiveness into question. In this paper, we investigate the impact of implementing such algorithms for congestion control in low voltage feeders. We review the latest NUM based algorithms for real-time EV charging control, and evaluate their behavior and impact on the comprehensive IEEE European Low Voltage Test Feeder. Our results show that the EV NUM problem can effectively capture the relevant operational constraints, as long as ampacity violations are the main bottleneck. Moreover, the results demonstrate an advantage of the primal NUM solution over the more conventional dual NUM solution in preventing a system overload.



\end{abstract}


%
\IEEEpeerreviewmaketitle

%
%

\section{Introduction} 

\IEEEPARstart{T}{he} Electric Vehicle (EV) numbers are expected to increase significantly in the coming years \cite{EVintegration}.
If EVs become prevalent, their charging will cause a huge load increase, which, without control, could destabilize distribution grids \cite{EVChargingProblem, putrus2009impact,richardson2010impact,gomez2003impact}.
The reason for this is that the power consumption of an EV charger can be up to 5-10 times the average power consumption of a common household \cite{schroeder2012economics, botsford2009fast}. 
The added load of charging a large number of EVs would exceed the capacity that the distribution grid was designed to support, i.e., 20-30\% above peak load \cite{willis2004power}. One solution to this problem is to upgrade the distribution grid in order to handle the load increase. This, however, comes at a high cost and would slow down EV adoption. A more efficient solution is to use Information and Communication Technology (ICT) to control EV charging. The challenge of this approach results from the large number of EVs and their unknown spatial distribution. 
Furthermore, the distribution grid's state is highly dynamic and, if renewable energy is present, difficult to predict.  
 Thus, there is growing interest in developing a distributed system for the control of EV charging, that can adapt quickly to the fast dynamics of the grid, allows for local decision making of the individual devices, and globally optimizes the operation of the entire system \cite{EnergyInformatics}.\\

This article considers the use of Network Utility Maximization (NUM) distributed optimization algorithms for \mbox{real-time} EV charging control. In our scenario it is assumed that EVs want to charge as fast as possible. Hence, the goal is to maximize the EVs charging rate while preventing an overload of the grid and a subsequent blackout. A blackout can only be avoided, if we are able to react within milliseconds, to prevent the triggering of protection devices (usually around 200 ms after an overload). The NUM formulation allows the definition of highly efficient distributed control algorithms. Nontheless, its simple network constraints do not include many of the physical and operational constraints of distribution networks. Therefore, a comprehensive evaluation of the effects that the implementation of such algorithms would have on a realistic low voltage feeder is required. \\
The use of the NUM formulation for real-time EV charging control was first proposed in \cite{ardakanian2012realtime, Ardakanian2013a}, where the dual decomposition approach is used to obtain a distributed optimization algorithm. However, evaluations in \cite{ardakanian2014real} showed that the dual solution algorithm suffers from stability and scalability issues. To address these issues, we proposed a novel solution based on primal decomposition and analyzed its dynamic behavior for real-time EV charging control in \cite{JoseRiveraCDC14,RiveraEenergy}. Nevertheless, an evaluation of the scalability of both dual and primal solutions is missing, as well as a comprehensive comparisson of both algorithms. Moreover, since the NUM formulation does not consider grid constraints, such as maximal voltage variations, a much needed evaluation of the impact of both algorithms on a realistic low voltage feeder is required. In this article, we extend our original work and provide a comprehensive comparison of the primal and dual EV NUM solution algorithms.  We focus on evaluating the scalability and reliability of both algorithms in static and dynamic scenarios. Most importantly, we evaluate the effect of implementing both algorithms for real-time EV charging control using comprehensive 3-phase power flow simulations on the IEEE European Low Voltage Test Feeder.  

This paper complements existing research in the area of real-time EV charging control in the following ways:
\begin{enumerate}
 \item We provide a coherent exposition of the EV NUM dual and primal decomposition approaches and outline their key characteristics with respect to actual deployment.
 \item We provide a comprehensive evaluation of the scalability and reliability of the different EV NUM solution algorithms for the IEEE European Low Voltage Test Feeder.
 \item We demonstrate with theoretical analysis and comprehensive experiments that the primal algorithm outperforms the dual algorithm in scalability and reliability. 
 \item We show that the NUM formulation can effectively be used to formulate EV charging control algorithms in low voltage feeders where ampacity constraints are the main bottleneck.
\end{enumerate}

The rest of the paper is organized as follows: In Section~\ref{section:ProblemFormulation}, we formulate the EV NUM problem. 
The primal and dual decomposition solutions are introduced in Section~\ref{sec:solution}. 
Our scalability and reliability evaluations based on the IEEE European Low Voltage Test Feeder are presented in Section~\ref{sec:evaluation} and a short discussion on the EV NUM formulation and the algorithms' applicability is carried out in Section \ref{sec:discussion}.  
Finally, Section~\ref{conclusion} provides our conclusions.

\section{Related Work}
The main challenge of real-time EV charging is the short time available to deliver a control response. 
The time range to avoid a fault in the distribution grid is usually in the hundreds of milliseconds \cite{willis2004power}. 
Some approaches to avoid the violation of grid constraints by controlling EV charging under real-time requirements have been proposed in the literature. 
In \cite{turitsyn2010robust}, an EV charging desynchronization approach to prevent too many charging EVs from overloading the grid is proposed. 
Similarly, in \cite{studli2012aimd}, an AIMD\footnote{The additive-increase/multiplicative-decrease (AIMD) algorithm is a feedback control algorithm best known for its use in TCP Congestion Avoidance.}-like control approach is proposed to avoid grid overloading. Although both approaches prevent grid overloading, they do not guarantee an optimal use of the available infrastructure. Hence, the power infrastructure is underutilized and the EVs' charging speed is reduced. In contrast to these approaches, we focus on the optimal use of the available infrastructure. We consider the use of optimization-based approaches and propose distributed optimization algorithms to cope with the real-time requirements.\\

Several studies have used distributed optimization techniques to design scalable EV charging control algorithms: 
In \cite{Gan.2012}, the valley filling problem is solved using a distributed subgradient approach. 
The authors of \cite{Mercurio2013} use the Alternating Direction Method of Multipliers (ADMM) for the load balancing of EVs at charging stations. 
In \cite{Ma2014}, the authors compare the use of several distributed optimization algorithms for the valley filling problem. 
We proposed a general framework that supports several EV aggregator control objectives in \cite{JoseRivera,RiveraEVADMM}. 
Nevertheless, all of the mentioned approaches are multiperiod optimizations, which in order to provide EV congestion control would require very accurate predictions of the EVs' location, charging requirements and the state of the grid. Such predictions are extremely challenging for a single low voltage feeder and can be highly inaccurate for individual EVs. An alternative, is to consider the solution of a single period optimization problem based on the current state of the system. The lack of prediction, however, requires very fast control responses to cope with the highly dynamic state changes of the power grid. This makes real-time EV charging a challenging time critical application. \\
A promising approach comes from the related problem of congestion control in communication networks and its implementation for real-time EV charging control is shown in \cite{Ardakanian2013a}, where the EV charging problem is formulated as a NUM, and dual decomposition is used to solve the problem. The NUM formulation defines an instantaneous problem. Hence, no prediction is required. Moreover, thanks to the simplicity of the NUM formulation the resulting algorithm is highly efficient. However, the proposed dual decomposition solution suffers from scalability problems, as the method can become unstable or in some situations may not adapt quickly enough to the grid dynamics \cite{ardakanian2014real}. To improve scalability and stability, we proposed a novel primal algorithm for the general NUM problem in \cite{JoseRiveraCDC14} and provided mathematical proof for its convergence to the optimal result under static conditions. Later in \cite{RiveraEenergy}, we derived a primal decomposition based real-time EV charging control algorithm and analyzed its behavior under dynamic conditions. Nevertheless, an analysis that compares the scalability of both algorithms and also evaluates their impact when implemented in a realistic distribution grid setting was missing. In this paper, we consolidate our previous work and contribute a comprehensive comparison of the dual and primal EV NUM  distributed optimization algorithms. We use the comprehensive IEEE European Low Voltage Test Feeder and conduct experiments on the behavior, scalability, and grid impact of both algorithms. This realistic distribution grid analysis is a significantly larger and more representative use case compared to the state-of-the-art. Since the NUM formulation does not consider losses and voltage constraints, the evaluation presented in this paper represents a vital contribution to NUM based real-time EV charging control.

\section{EV Network Utility Maximization (NUM)} 
\label{section:ProblemFormulation}

We formulate the real-time EV charging control problem as a Network Utility Maximization (NUM) problem (cf.~\cite{Kelly98}), where the objective is to maximize the utility of the network users while respecting the users' and the network's constraints. In our application, the users are the EV chargers and the network constraints are set by the maximal available loading of the grid devices.
In contrast to the standard NUM problem, our formulation has a maximal bound on the users' rate that represents the maximum charging rate of the EVs.

The EV NUM problem considers a distribution network composed of grid devices represented by a set $\mathcal{P} = \{1,\ldots,M\}$ of directed links with a finite capacity given by the vector $c=[c_l]_{l\in\mathcal{P}}$. This capacity is the maximum possible loading, i.e., the maximum loading minus the current load caused by all uncontrollable loads. 
The EV chargers are represented by set $\mathcal{V} = \{1,\ldots,N\}$. 
Our optimization variables are the EV charging rates denoted by the non-negative charging vector $x=[x_i]_{i\in \mathcal{V}}$. 
The maximum charging rate of the EV chargers are represented by the vector $\overline{x}=[\overline{x}_i]_{i\in \mathcal{V}}$. 
The energy flow to each EV charger traverses several grid devices before reaching its destination. 
We define this as the \textit{route} to the EV charger. 
The capacity constraints at the grid devices can be expressed in vector form as $Rx \leq c$, where R is the \textit{routing matrix} of dimension $M \times N$:
\begin{equation}
R_{li}= \left\{ 
  \begin{array}{l l}
    1 & \quad \text{if $l$ is on the route to $i$}\\
    0 & \quad \text{otherwise.}
  \end{array} \right.
\end{equation}

For each EV charger $i$, we consider the utility function $U_i(x_i)=w_i\log(x_i)$, where $w_i > 0$ is a weight parameter that can be used to set priorities, e.g., to allow some EVs to always charge at higher rates than others. The weighted logarithm utility function is used in order to obtain a proportionally fair allocation of the available resources \cite{mo2000fair}, i.e., the logarithmic utility function assures that each EV charger receives at least a minimal amount of charging rate, because if $x_i=0$, then the utility of the EV is $U(0)=-\infty$. With these definitions, the EV NUM problem can be formulated as follows:
\begin{Problem} \label{for:problem}
 The EV NUM problem \\
 \begin{equation}
 \begin{array}{ll}
 \underset{x}{\text{minimize}} &\sum_{i=1}^{N}- w_i \log(x_i) \\
 \text{subject to}    &R x \leq c\\
   & 0 \leq x \leq \overline{x}  
  \end{array}
 \end{equation}
\end{Problem}

\section{Distributed EV NUM solution} \label{sec:solution}

In the following we use dual decomposition and primal decomposition to formulate distributed optimization algorithms that solve the EV NUM problem. 

\subsection{Dual decomposition solution}

To formulate the dual decomposition algorithm we first define EV NUM's Lagrangian:
\begin{eqnarray}
 L(x,\lambda)=  \sum_{i=1}^N - w_i log(x_i) + \lambda^T (R x - c) , \label{for:lagrangian}
\end{eqnarray}
where $\lambda=[\lambda_l]_{l\in\mathcal{P}}$ is the vector of Lagrangian multipliers, also known as dual variables. 
Then, we decompose the optimization problem into two sub-problems, of which the first optimizes the Lagrangian function with respect to the primal variables $x$, and the second one with respect to the dual variables $\lambda$.
The sub-problems are linked via mutual updates of the $x$ and $\lambda$ variables in an iterative process.
The resulting dual algorithm is:
\begin{eqnarray}
x^{k+1} &=& \underset{0 \leq x \leq \overline{x} }{\mbox{argmin}}\  L(x,\lambda^k) \label{for:xProblem} \\
\lambda^{k+1} &=& \underset{\lambda \geq 0}{\mbox{argmax}}\  L(x^{k+1},\lambda) \label{for:lambdaProblem},
\end{eqnarray}
where $k$ is the iteration index. 
The $0 \leq x \leq \overline{x}$ condition in the $x$-update (\ref{for:xProblem}) results from the original EV NUM problem definition and the $\lambda \geq 0$ condition in the $\lambda$-update (\ref{for:lambdaProblem}) results from the Karush-Kuhn-Tucker (KKT) conditions. 
The $x$-update (\ref{for:xProblem}) has an analytic solution, which can be obtained by setting the partial derivative of (\ref{for:lagrangian}) with respect to $x_i$ to zero, solving for $x_i$, and then projecting this value onto $0\leq x_i\leq \overline{x}_i$. 
The $\lambda$-update has no analytic solution, thus, a gradient projection method to approximate the solution is used. 
The result is a dual decomposition algorithm solving the NUM problem, which can be written for each element of $x$ and $\lambda$ as follows:
\begin{eqnarray}
 x_i^{k+1} &=& \min \left\{ \frac{w_i}{R_i^T \lambda^k},  \overline{x}_i \right\}, \label{for:dualxupdate}   \\
\lambda_l^{k+1} &= & \max \left\{ \lambda_l^k + \kappa ( R_l x - c_l),0\right\},  
\end{eqnarray}
where $R_i$ is the $i$-th column and $R_l$ the $l$-th row of routing matrix $R$ and $\kappa \geq 0$ is the step size of the gradient projection method. The dual decomposition solution is summarized in Algorithm \ref{algorithm_dual}. 
The EV NUM dual algorithm only guarantees satisfaction of constraints $Rx<c$ upon convergence.

\begin{algorithm}[h] 
\SetKwInOut{Input}{input}\SetKwInOut{Output}{output}
  \caption{EV NUM dual decomposition algorithm}
  \label{algorithm_dual}
  $k=0$
  
  \While{true}{

 
\lnlset{InResR1}{1)}EV charger updates charging rate \\
 \For{$i=\{1,\ldots,N\}$}{  
  $x_i^{k+1}= \min \left\{ \frac{w_i}{R_i^T \lambda^k},  \overline{x}_i \right\}$ }

\lnlset{InResR2}{2)}Network devices update price\\
 \For{$l=\{1,\ldots,M\}$}{
 $\lambda_l^{k+1}= \max \left\{ \lambda_l^k + \kappa ( R_l x - c_l),0\right\}$}

\lnlset{InResR3}{3)}Check for convergence\\
  	\If{$||\lambda^{k+1} - \lambda^k||_2 \leq \epsilon_{d}$}
  	{	\textbf{break}}

  
  	{$k=k+1$}
  	
  	}
  	
\end{algorithm}

\subsubsection{Convergence}  
\label{sec:convergence_dual}

The convergence criteria for dual decomposition is given by the following condition: 
\begin{equation}
||\lambda^{k+1} - \lambda^k||_2 \leq \epsilon_{d},
\end{equation}
where $\epsilon_{d}>0$ is the convergence parameter.
According to \cite{Ardakanian2013a} the convergence of the algorithm is guaranteed if
\begin{equation}
 0 < \kappa \leq \frac{2}{\overline{x}_{max} \overline{L} \overline{N} }, \label{for:incentiveConvergence}
\end{equation}
where $\overline{x}_{max} = \max{\overline{x_i}}$ is the maximum charging rate across all EV chargers, $\overline{L}=\max_i \sum_l R_{li}$ is the maximum number of constrained network devices being used by any EV charger, and $\overline{N}=\max_l R_{li}$ is equivalent to the maximum number of EV chargers using any network device. 
The condition in Eq. \ref{for:incentiveConvergence} defines a theoretical upper bound for the step size value. 
Beyond this value, stability is not guaranteed, and the algorithm could become unstable with oscillating EV charging rates.
%

\subsubsection{Interpretation} 

The dual decomposition algorithm can be seen as an incentive-based control approach. 
In Algorithm \ref{algorithm_dual}, the dual variables $\lambda$ can be regarded as prices that the network devices define to users for the use of the available resources. 
These prices in turn influence the charging rate of the EVs. 
Hence, the algorithm iteratively modifies prices based on the users response and converges towards an optimal price that maximizes the EVs' utility without violating grid constraints. 

\subsection{Primal decomposition solution}

We now consider a primal decomposition solution algorithm. 
The advantage of this approach is that it results in an algorithm that has the anytime property, i.e., the results on each iteration are feasible. 
To formulate a primal decomposition solution, we first modify the original problem to include an upper bound for each EV charger, which we call budget.
With $b=[b_i]_{i\in \mathcal{V}}$ as the vector of the EV chargers' budgets, we introduce a new set of primal variables and reformulate the problem as follows:
\begin{Problem} \label{for:problemBudgets}
EV NUM with budgets
\begin{equation}
\begin{array}{ll}
\underset{x,b}{\mbox{minimize}} & \sum_{i=1}^{N} -w_ilog(x_i)  \\
\text{subject to}     &       x \leq b \\
      &   R  b \leq c \\
      & 0 \leq x \leq\ \overline{x}  \\
      \end{array}
\end{equation}
\end{Problem}

In \cite{JoseRiveraCDC14} we provided a mathematical proof, which confirms that the solution of the EV NUM problem with budgets is equivalent to the solution of the original EV NUM problem. 
Hence, we can implement primal decomposition on the EV NUM problem with budgets to obtain a distributed primal algorithm that solves the original EV NUM problem. 
We define the Lagrangian of Problem \ref{for:problemBudgets} as follows:
\begin{equation}
 L(x,b,\mu) = \sum_{i=1}^{N} -w_ilog(x_i)  + \mu (x-b), \label{for:primal_lagrangian}
\end{equation}
where $\mu=[\mu_i]_{i\in \mathcal{V}}$ is the vector of the Lagrangian variables. 
The primal decomposition results in a formulation consisting of two sub-problems. 
The first one optimizes over a set of the primal variables $x$, the second one over the other set of the primal variables $b$.
The resulting primal algorithm is as follows:
\begin{eqnarray}
 (x^{k+1}, \mu^{k+1}) &=& \underset{\mu \geq 0}{ \mbox{argmax} }\ \underset{0 \leq x \leq\ \overline{x}}{\mbox{argmin}}\  L(x,b^k,\mu) \label{for:xproblem_primal} \\
 b^{k+1} &=& \underset{R  b \leq c}{\mbox{argmin}}\ L(x^{k+1},b,\mu^{k+1}) \label{for:bproblem_primal}
\end{eqnarray}

In the $x$-update in Eq. \ref{for:xproblem_primal}, the  $\mu \geq 0$ conditions result from the KKT conditions of the dual variables, and the $0 \leq x \leq\ \overline{x}$ conditions result from the original EV NUM problem. 
In the $b$-update provided in Eq. \ref{for:bproblem_primal}, the $R  b \leq c$ conditions result from the EV NUM problem with budgets.  

The x-update in Eq. \ref{for:xproblem_primal} has an analytic solution, which can be written for each of its elements as
\begin{eqnarray}
x_i^{k+1}&=&\min\{ b_i^k,\overline{x}_i\}, \label{for:EVupdate1}\\
 \mu_i^{k+1}&=&\left\{ 
  \begin{array}{l l}
    0 & \quad \text{if $x_i^{k+1}=\overline{x}$}\\
    \frac{w_i}{x_i^{k+1}} & \quad \text{otherwise}.
  \end{array} \right.\label{for:EVmarginalBenefit}
\end{eqnarray}
The $b$-update provided by Eq. \ref{for:bproblem_primal} can be solved using a gradient projection method as explained in \cite{BertsekasDCBook}. 
However, this approach would require the current available loading $c_l$ of all network devices to be sent to a central location in each iteration. 
In a large distribution grid, this would cause significant communication overhead. 
Therefore, we use the \textit{sequential projections method} described in \cite{LectureBoyd} together with gradient descent, which results in a \textit{gradient sequential projection method}. 
In our approach, we first update the budgets using the gradient descent method and thereafter sequentially project the updated budgets onto the constraint of each network device. 
The sequential projections method for the $b$-update can be written as follows:
\begin{equation}
b^{k+1}= \mathbf{P}_{\mathcal{C}_M} \{\ldots \mathbf{P}_{\mathcal{C}_2} \{ \mathbf{P}_{\mathcal{C}_1} \{  b^k + \gamma \mu^{k+1}  \}\}\ldots\}, \label{for:procedure}
\end{equation}
where $\gamma>0$ is the gradient step size, $\mathbf{P}$ is a projection operator, and $\mathcal{C}_l$ is the constraint set defined as:
\begin{equation}
\mathcal{C}_l=\{ b |\  R_l  b \leq c_l \}, \label{for:procedure2}
\end{equation}
where $R_l$ is the $l$-th row of the routing matrix $R$.


The projections defined in Eq. \ref{for:procedure} are all projections onto a halfplane, which have an analytic solution \cite{BoydBook}. 
Hence, in (\ref{for:procedure}) each constrained network device $l=1,\ldots,M$, modifies the budgets as follows:
\begin{equation}
b^{k+1}= \left\{ 
  \begin{array}{ll}
    b^{k+1}, & \text{if $R_l b^{k+1} \leq c_l$}\\
    b^{k+1} +  (c_l - R_l b^{k+1})R_l^T/||R_l||_2^2 ,  & \text{otherwise}.
  \end{array} \right. \label{for:protectionProjection}  
\end{equation}

The resulting primal decomposition solution is summarized in Algorithm \ref{algorithm_primal}. 
The EV NUM primal algorithm guarantees that the resulting charging rates for EVs are feasible on each iteration, i.e., $Rx<c$ for all $k$.

\begin{algorithm}[t] 
\SetKwInOut{Input}{input}\SetKwInOut{Output}{output}
  \caption{EV NUM primal decomposition algorithm}
  \label{algorithm_primal}
  $k=0$
  
  \While{true}{

 
\lnlset{InResR1}{1)}EV charger updates charging rate \\
 \For{$i=\{1,\ldots,N\}$}{  
  $x_i^{k+1}= \min\{ b_i^k,\overline{x}_i\}$

  \eIf{$x_i^{k+1}=\overline{x}_i$}
  {$\mu_i^{k+1}= 0$}
  { $\mu_i^{k+1}=\frac{w_i}{x_i^{k+1}}$}
}  

\lnlset{InResR2}{2)}Network devices update budgets\\
 $b^{k+1}= b^k + \gamma \mu^{k+1}$

\lnlset{InResR3}{3)}Check for convergence\\
  	\If{$||b^{k+1} - b^k||_2 \leq \epsilon_p$}
  	{	\textbf{break}}

\lnlset{InResR4}{4)}Budgets are projected to network constraints	
  
  \For{$l=\{1,\ldots,M\}$}{
  \If{$R_l b^{k+1} > c_l$  }{\vspace{0.1cm}
$b^{k+1} =  b^{k+1} +  (c_l - R_l b^{k+1})R_l^T/||R_l||_2^2$}
  }
  	{$k=k+1$}
  	
  	}
  	
\end{algorithm}


\subsubsection{Convergence}
\label{sec:convergence_primal} 

The convergence criteria for the EV NUM primal algorithm is given by the following threshold condition:
\begin{equation}
 ||b^{k+1} - b^k||_2 \leq \epsilon_p,
\end{equation}
where $\epsilon_p>0$ is the convergence threshold.

In \cite{JoseRiveraCDC14} we provide a mathematical proof for the converge of the algorithm to within a distance of $K^2\gamma/2$ from the optimal solution, where $K$ is the Lipschitz constant of the value function of the b-update problem in Eq. \ref{for:bproblem_primal} and $\gamma$ is the algorithm step size. 
This result reveals that the algorithm's stability does not depend on step size.

\subsubsection{Interpretation} 
\label{sec:interpretation_primal} 

The primal decomposition algorithm can be seen as a budget-based control approach. 
In Algorithm \ref{algorithm_primal}, the network devices define budgets $b$ as upper bounds for the charging rate of the EVs. 
The EVs in turn report their marginal benefits $\mu$ to a single centralized location that updates the budgets. 
The primal algorithm updates the budgets based on the users' marginal benefit and also sequentially projects the updated budgets onto the network devices contraints in order to gurantee fesibility. 
This iterative process converges to the optimal budget values that maximize the EVs' utility without violating the grid constraints. 
The primal distributed solution guarantees that the constraints of the EV NUM problem are feasible on each iteration, such that the produced control values can be used on each iteration without the risk of overloading the system.


\section{Evaluation} 
\label{sec:evaluation}

We have conducted three experiments to evaluate the dual and primal EV NUM solution algorithms. 
The first experiment evaluates the scalability and convergence of both algorithms under static conditions.
The second experiment looks at the behavior of both algorithms under dynamic conditions.
Finally, the third experiment shows the impact that both algorithms have on the voltages and currents of a distribution grid. 
The source code and data of all our experiments can be found online \footnote{\href{https://github.com/chepeadan/EVNUM}{github.com/chepeadan/EVNUM}}.

\begin{figure}[t!]
\centering
\vspace{-0.25cm}
\includegraphics[width=.9\columnwidth]{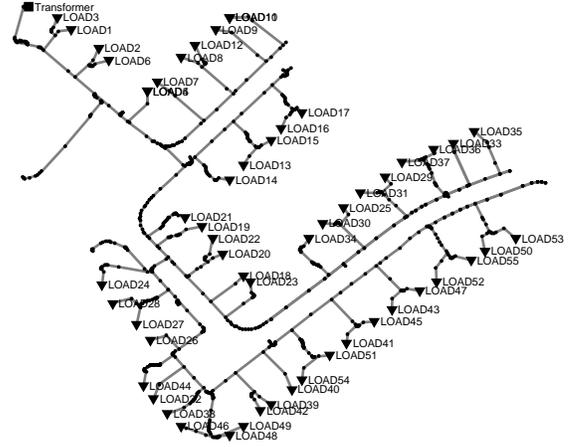}
\vspace{-0.25cm}
\caption{IEEE European Low Voltage Test Feeder.}
\label{fig:grid}
\end{figure}

All our experiments are based on the IEEE European Low Voltage Test Feeder \cite{EULVgrid}. 
Our evaluation grid, shown in Fig. \ref{fig:grid}, is a three-phase radial distribution feeder at the voltage level of 416 V (phase-to-phase) with a total of 906 buses, 905 lines, and 55 loads. 
All relevant data for our experiments are available on the test case data, with the exception of the lines' ampacity. 
Thus, we define the ampacity based on empirical data of similar standard test networks \cite{IEEERadialTestFeeders}. 
The ampacity values used in our evaluations are summarized in Table \ref{table:ampacity}.
We assume that each load has one EV charger with a maximal charging capacity of 20 kW (3-phase). 
We also assume that the EVs start arriving at 5 p.m. according to a Poisson distribution with arrival rate of 1 per minute. 
Furthermore, the EVs are assumed to be fully discharged upon arrival and wishing to be fully charged to their maximal capacity of 24 kWh.

To formulate the EV NUM optimization problem, we assume constant voltages. 
Hence, our optimization variables $x$ are the currents that the EV chargers draw from the network. 
The maximum available capacity of the network devices $c$ is defined by the lines' ampacity minus the current drawn by the loads. 
The maximum charging rate $\overline{x}$ is given by the maximum charging current of the EV chargers. 
Without loss of generality, we assume that all EVs have the same importance, i.e., $w_i = 1$.

To make use of the EV NUM formulation, one needs to assume a constant voltage, because if voltage is constant, then the main network constraints are the line ampacity limits, which are linear and can be expressed with the NUM formulation $Rx< c$, see also \cite{JoseRiveraCDC14,RiveraEenergy,ardakanian2012realtime,Ardakanian2013a,ardakanian2014real}. While omitting voltage constraints is risky, as our evaluations will show, the NUM model offers a good approximation, when ampacity violations happen before voltage violations. In such cases, the NUM formulation offers a good trade-off between model accuracy and the required simplicity to design distributed algorithms that can respond in the millisecond time scale. 
%

\begin{table}[t]
\centering
\caption{Power line ampacity used in evaluation.}\vspace{.25cm}
\label{table:ampacity}
\begin{tabular}{|l|c|} \hline
\textbf{Line code name}  & \textbf{Ampacity (A)} \\ \hline
2c\_.007			& 56 \\ \hline
2c\_.0225		& 83 \\ \hline
2c\_16			& 83 \\ \hline
35\_SAC\_XSC		& 110 \\ \hline
4c\_.06			& 210 \\ \hline
4c\_.1			& 560 \\ \hline
4c\_.35			& 210 \\ \hline
4c\_185			& 405 \\ \hline
4c\_70			& 560 \\ \hline
4c\_95\_SAC\_XC		& 180\\ \hline
\end{tabular}
\end{table}

\subsection{Static evaluation} 
\label{sec:static_evaluation}

We evaluate the convergence of the dual and primal EV NUM algorithm for different step size values and consider the scalability behavior of both algorithms for different numbers of EVs and varying grid size. 
The static evaluation is characterized by constant optimization parameters, i.e., the parameters of the EV NUM problem don't change over time. 
The static case is the result of the fixed loads having a constant value, which leads to a constant maximal available loading of the devices $c$. 
Without loss of generality, we assume that the network is single phase for the static experiments. 
Hence, the size parameters for our EV NUM problem are $N=55$ and $M=905$. 
A centralized solution of the EV NUM problem is used as a reference for the optimal result.

\begin{figure*}[!t]
  \centering
\begin{subfigure}{.49\textwidth}
  \centering
  \includegraphics[width=\linewidth]{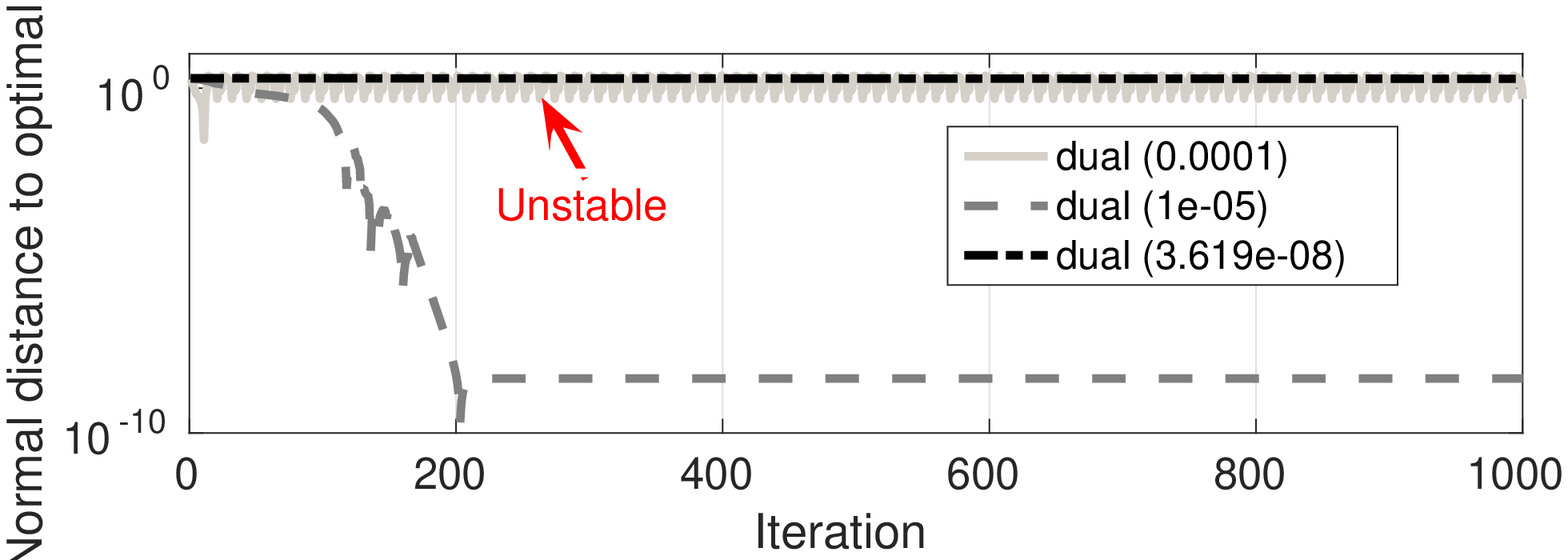}
  \caption{Different step size}
  \label{fig:dual_convergence}
\end{subfigure}%
\begin{subfigure}{.24\textwidth}
  \centering
  \includegraphics[width=1\linewidth]{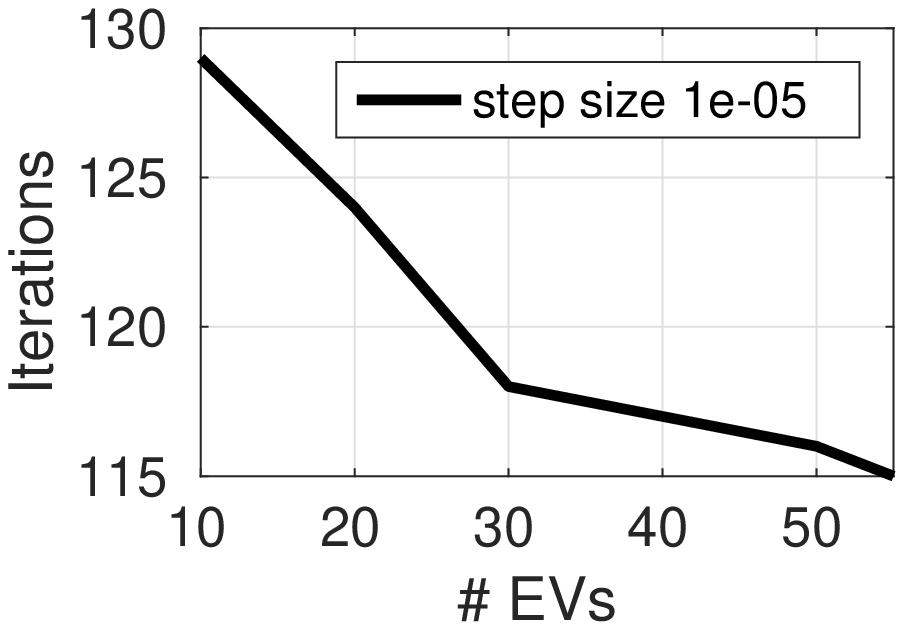}
  \caption{Different number of EVs}
  \label{fig:dual_convergence_N}
\end{subfigure}
\begin{subfigure}{.24\textwidth}
  \centering
  \includegraphics[width=1\linewidth]{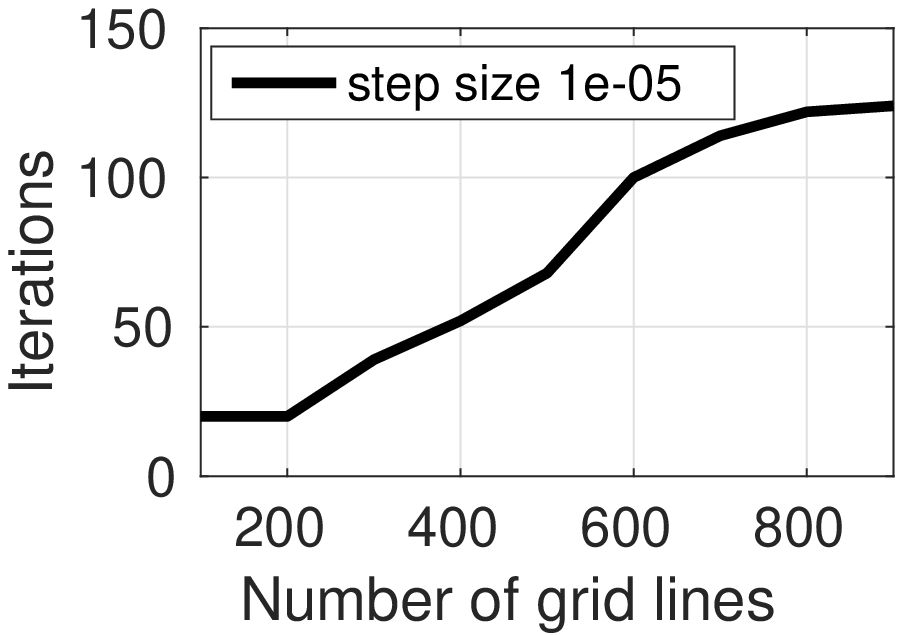}
  \caption{Different grid size}
  \label{fig:dual_convergence_M}
\end{subfigure}
\caption{EV NUM dual algorithm static behavior (each iteration $\sim$ 20 ms)}
\label{fig:static_dual}
\end{figure*}

The convergence experiment results for the dual algorithm in Fig \ref{fig:dual_convergence} show that the dual algorithm becomes unstable when the step size is too large ($\kappa = 0.0001$). 
To avoid instability, a theoretical upper bound for the step size value was defined in Section \ref{sec:convergence_dual}, which guarantees dual algorithm convergence ($\kappa = 3.619e-8$). 
Nonetheless, the theoretical upper bound is usually too conservative and a stable and faster step size can be used, e.g., $\kappa = 1e-05$.

\begin{figure*}[!t]
  \centering
\begin{subfigure}{.49\textwidth}
  \centering
  \includegraphics[width=\linewidth]{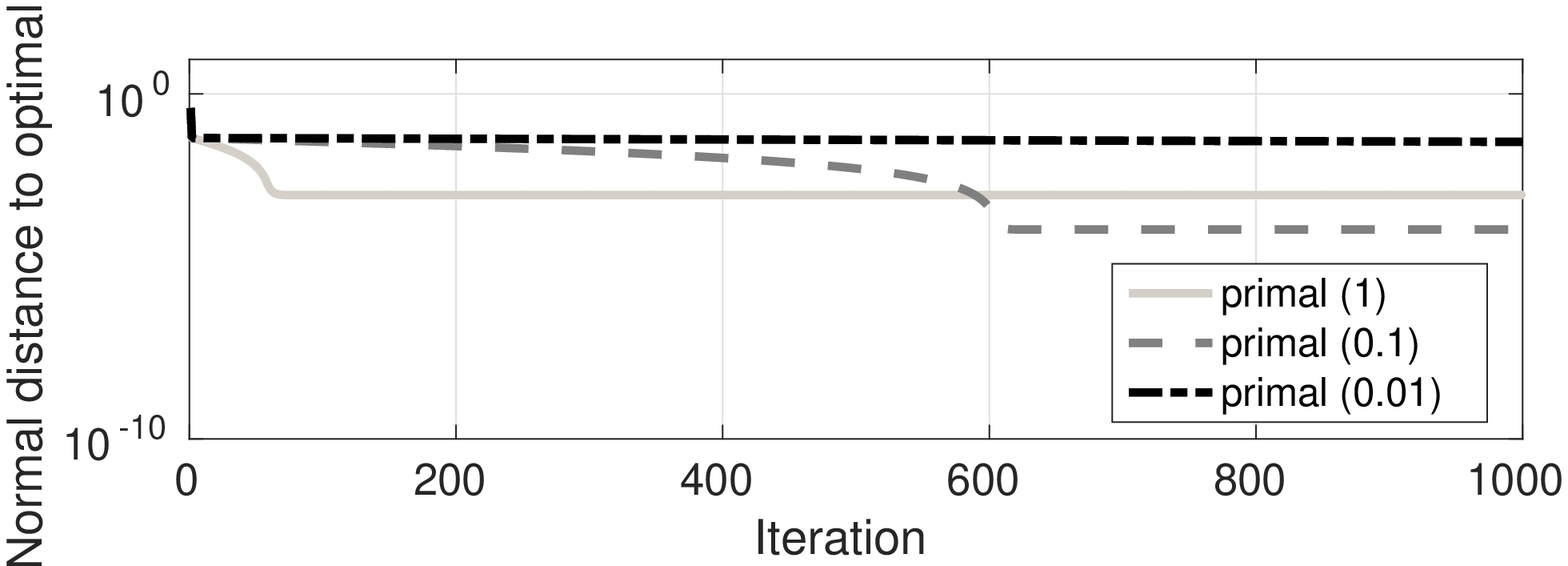}
  \caption{Different step size}
  \label{fig:primal_convergence}
\end{subfigure}%
\begin{subfigure}{.24\textwidth}
  \centering
  \includegraphics[width=1\linewidth]{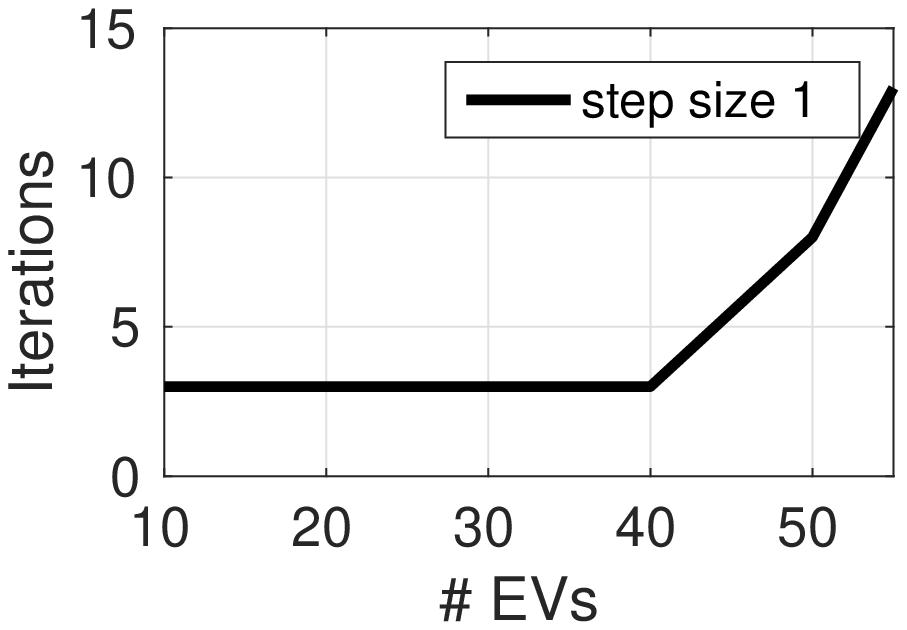}
  \caption{Different number of EVs}
  \label{fig:primal_convergence_N}
\end{subfigure}
\begin{subfigure}{.24\textwidth}
  \centering
  \includegraphics[width=1\linewidth]{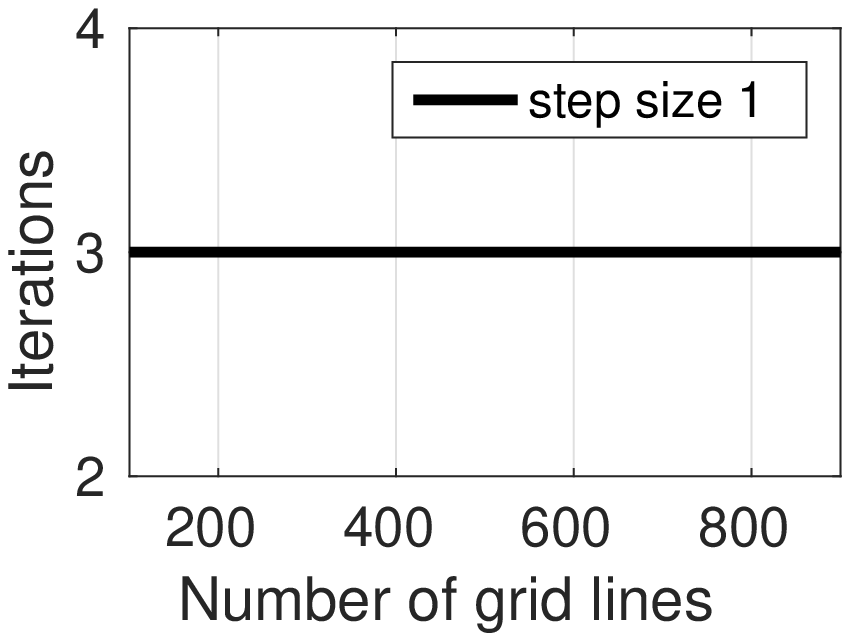}
  \caption{Different grid size}
  \label{fig:primal_convergence_M}
\end{subfigure}
\caption{EV NUM primal algorithm static behavior (each iteration $\sim$ 20 ms)}
\label{fig:static_primal}
\end{figure*}

The convergence results for the primal algorithm in Fig. \ref{fig:primal_convergence} show that the primal algorithm does not become unstable and converges within an increasing distance to the optimum with increasing step size. 
Hence, our experimental results for the primal algorithm confirm the theoretical behavior discussed in Section \ref{sec:convergence_primal} and demonstrate that the primal algorithm does not suffer from the instability issues of the dual algorithm.

To compare the scalability of both algorithms, we look at their convergence behavior as the EV NUM problem size parameters vary and measure the number of iterations required to reach 95\% convergence. 
First, we modify $N$ from 10 to 50 in increments of 10, which is equivalent to changing the number of EVs in the network. 
Then, we modify $M$ from 100 to 900 in increments of 100, which is equivalent to modifying the size of the network. The results in Fig \ref{fig:dual_convergence_N} and Fig \ref{fig:dual_convergence_M} reveal that the dual algorithm is highly sensitive to changes in the number of EVs and grid size. In contrast, the results for the primal algorithm in Fig. \ref{fig:primal_convergence_N} and Fig. \ref{fig:primal_convergence_M} show that its convergence behavior remains almost constant for changes in the problem size parameters. Hence, our experiments reveal that the primal algorithm is less sensitive to variations in problem size and therefore offers significant scalability advantages over the dual algorithm. Moreover, unlike the primal algorithm, the dual algorithm does not guarantee feasible control values on each iteration, which increases the chance of a blackout. While the computation time of both algorithms can be neglected (close form solutions), both require communication on each iteration. If we assume a maximal communication delay of 20 ms per iteration and a minimal protection tripping time of 200 ms, then to guarantee a feasible control values, the dual algorithm would need to converge in less than 10 iterations, which is well bellow the actual number of iterations required by the dual algorithm in our results.

\subsection{Dynamic evaluation} 
\label{sec:dynamic_evaluation}

In the dynamic evaluation, we consider the load dynamics, which cause the maximal available network load $c$ to change over time. 
We make use of the load shapes included in the IEEE European Low Voltage Test Feeder data, which are time series with a one-minute time resolution over 24 hours. 
The dual and primal algorithm are implemented in one-minute resolution, i.e., on each iteration of the algorithms the maximal available load of the devices $c$ change. 
We use a step size of 0.00001 for the dual algorithm and 1 for the primal algorithm. 
The problem size parameters for our three-phase evaluation network are $N=55$ and $M=3 \cdot 905$.

\begin{figure*}[!t]
  \centering
\begin{subfigure}{.48\textwidth}
  \centering
  \includegraphics[width=\linewidth]{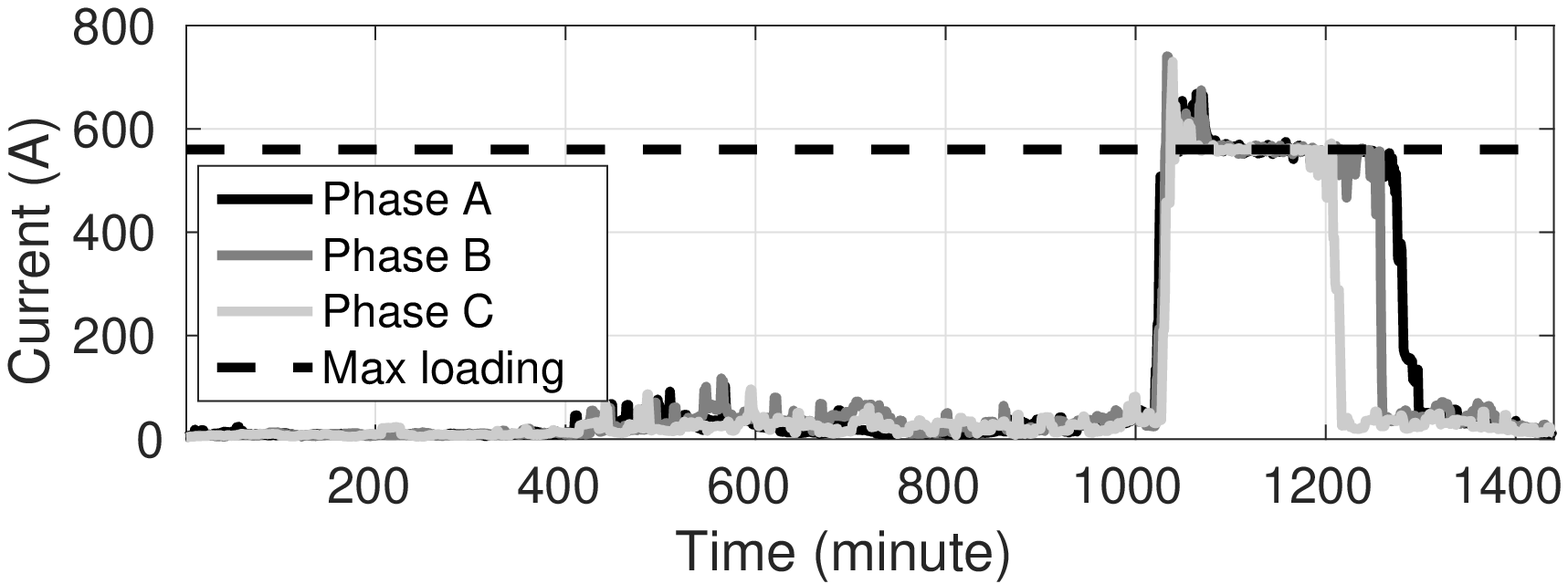}
  \caption{Dual algorithm result for the main power line}
  \label{fig:dual_result}
\end{subfigure}
\begin{subfigure}{.48\textwidth}
  \centering
  \includegraphics[width=1\linewidth]{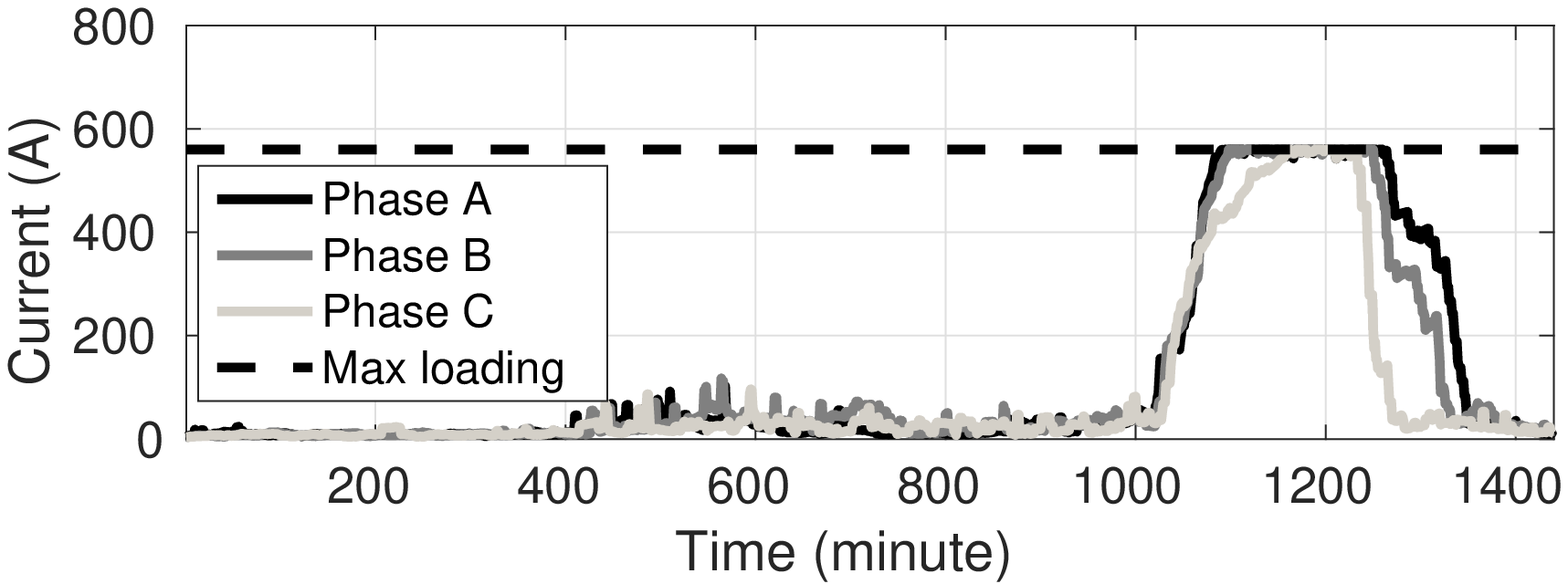}
  \caption{Primal algorithm result for the main power line}
  \label{fig:primal_result}
\end{subfigure}
\caption{EV NUM dynamic behavior.}
\label{fig:dynamic_NUM}
\end{figure*}

Fig. \ref{fig:dynamic_NUM} shows the results of the EV NUM optimization with the dual and the primal algorithm for the main power line. 
The result for the dual algorithm in Fig. \ref{fig:dual_result} shows that the maximum load condition of the line is violated as soon as EVs start to arrive. 
This result is expected, since the dual algorithm does not offer any guarantees that the network constraints $Rx<c$ will be fulfilled during runtime. 
To avoid overloading of the devices, the dual algorithm needs to be given enough time to come close to convergence. 
Hence, the dual control algorithm would need to be implemented at a higher frequency.

The result for the primal algorithm in Fig. \ref{fig:primal_result} shows the desired behavior for real-time EV control: 
The algorithm makes maximum use of the network without overloading it. 
As explained in Section \ref{sec:interpretation_primal}, the primal algorithm has the anytime property, which guarantees that the constraints of the EV NUM problem are fulfilled on each iteration. 
The anytime property gives the primal algorithm an advantage over the dual algorithm, because we don't need to provide the algorithm with enough iterations in order to guarantee control values that respect the network constraints. 
Hence, the primal algorithm can be implemented at a lower frequency than the dual algorithm. 

\subsection{Effects on a distribution network} 
\label{sec:evaluation_effect}

\begin{figure*}[!t]
  \centering
\begin{subfigure}{.49\textwidth}
  \centering
\includegraphics[width=\linewidth]{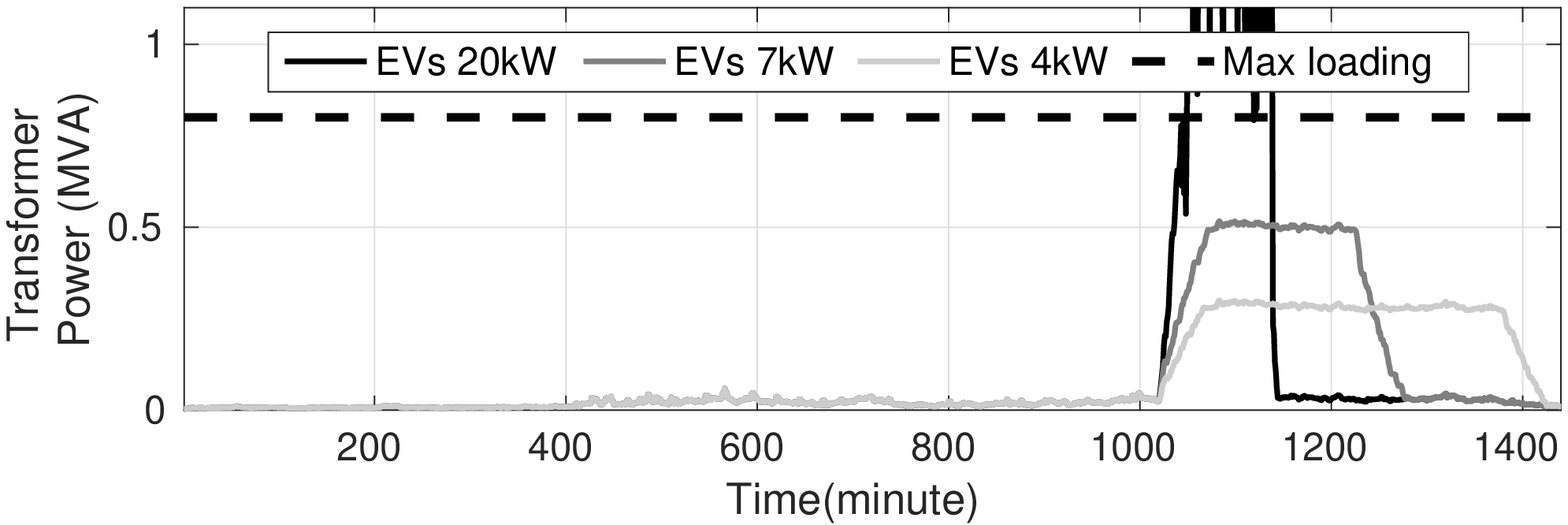}
\includegraphics[width=\linewidth]{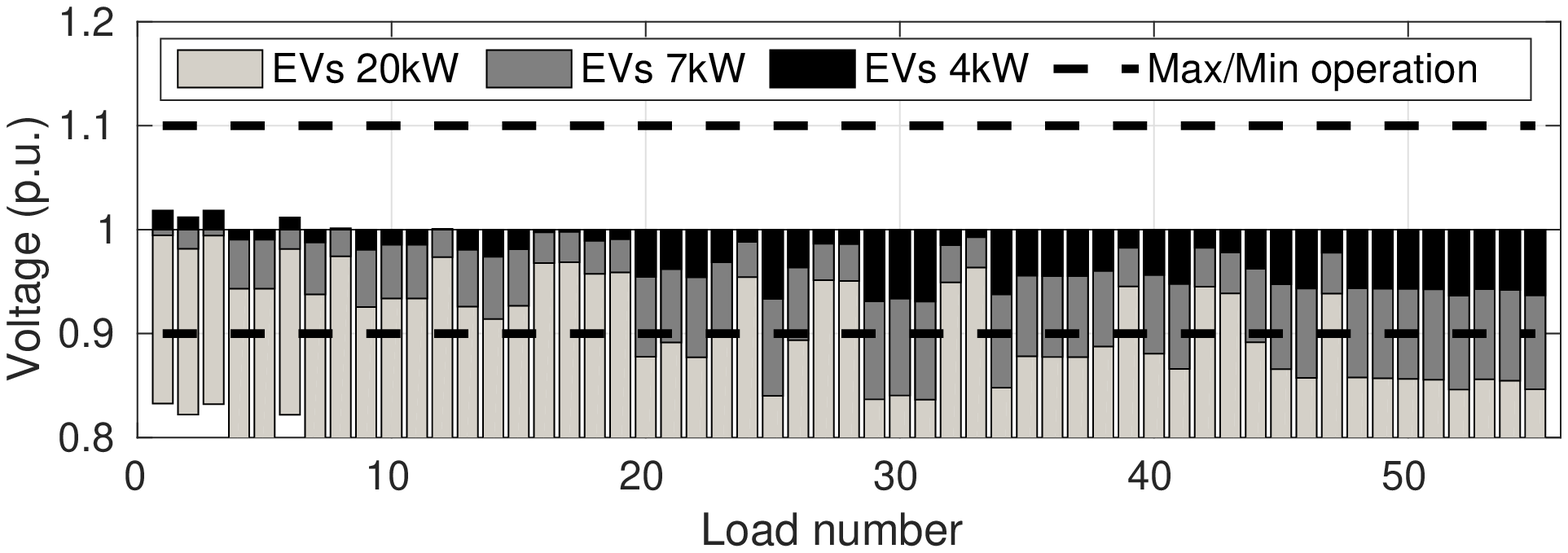}
\includegraphics[width=\linewidth]{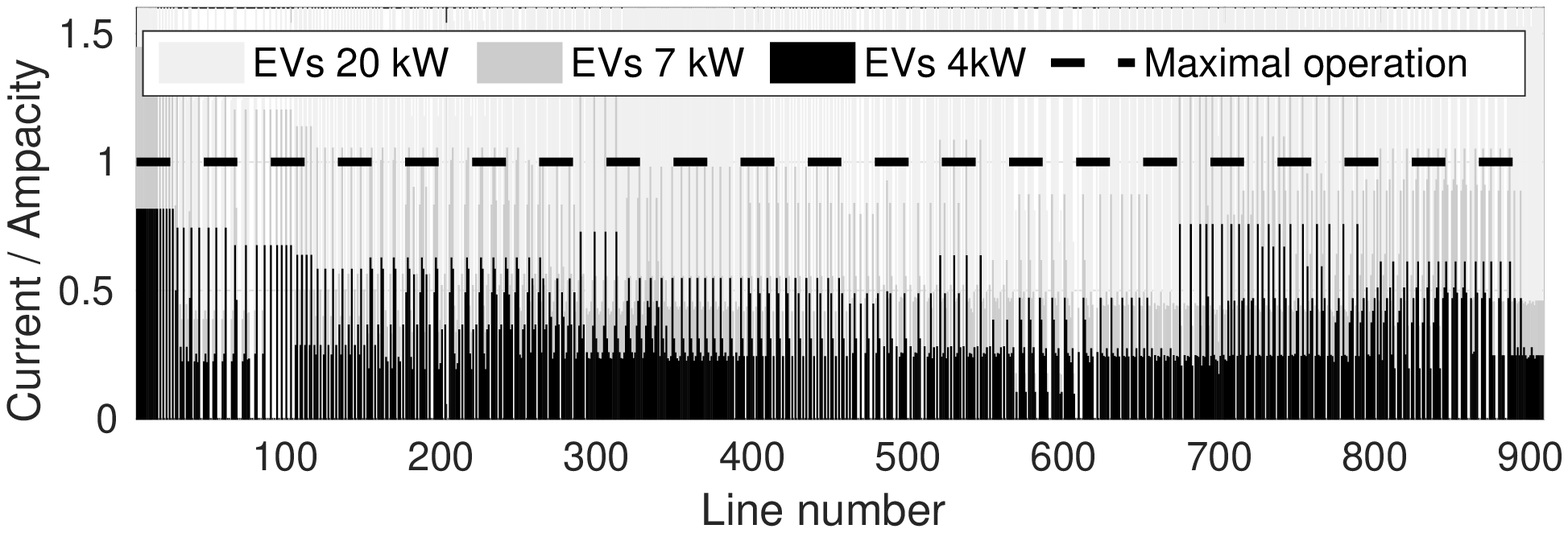}
  \caption{No control (voltages and currents at peak time).}
  \label{fig:no_control}
\end{subfigure}
\begin{subfigure}{.49\textwidth}
  \centering
\includegraphics[width=\columnwidth]{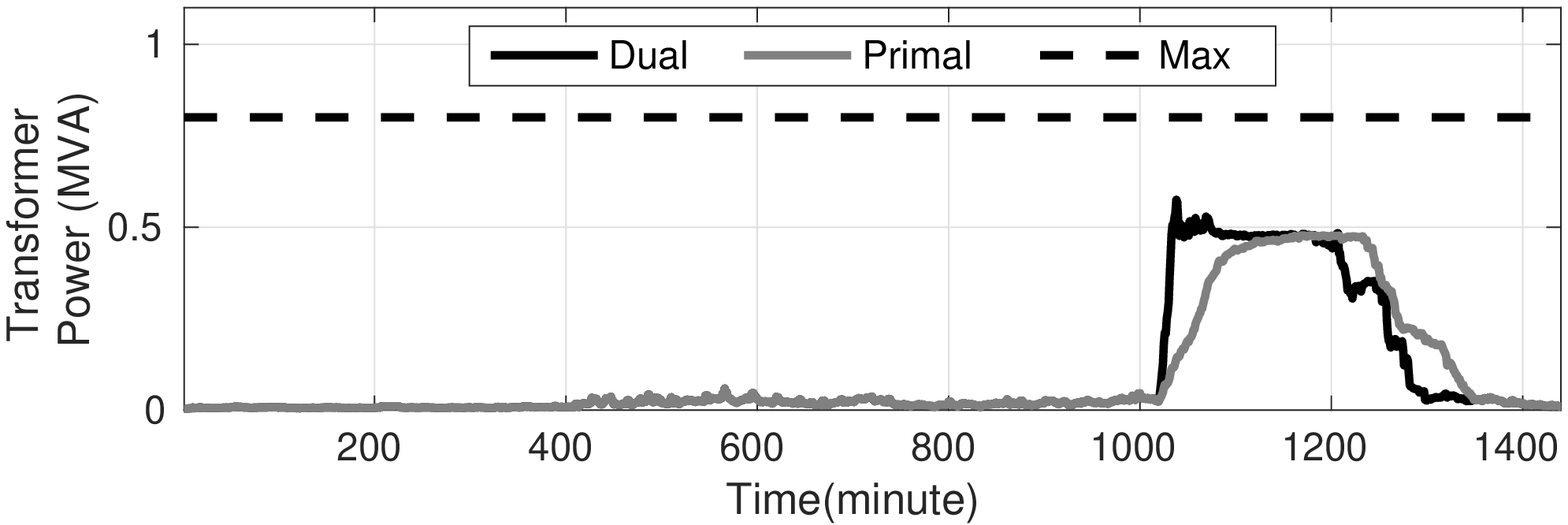}
\includegraphics[width=\columnwidth]{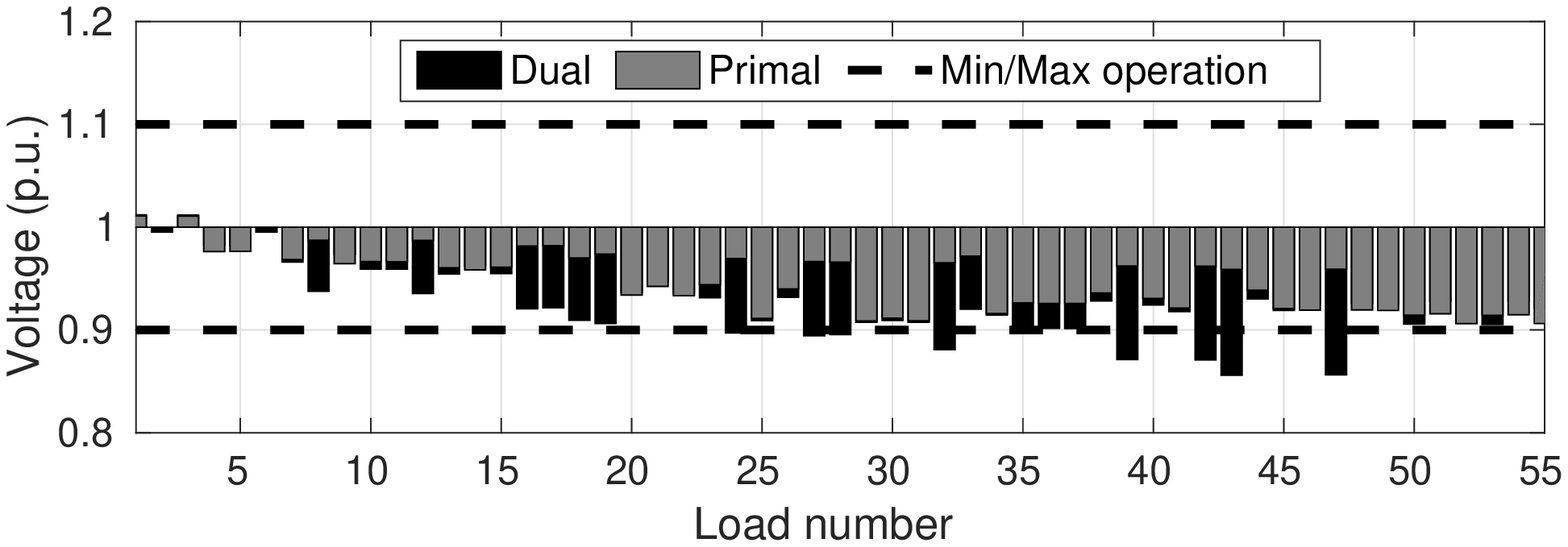}
\includegraphics[width=\columnwidth]{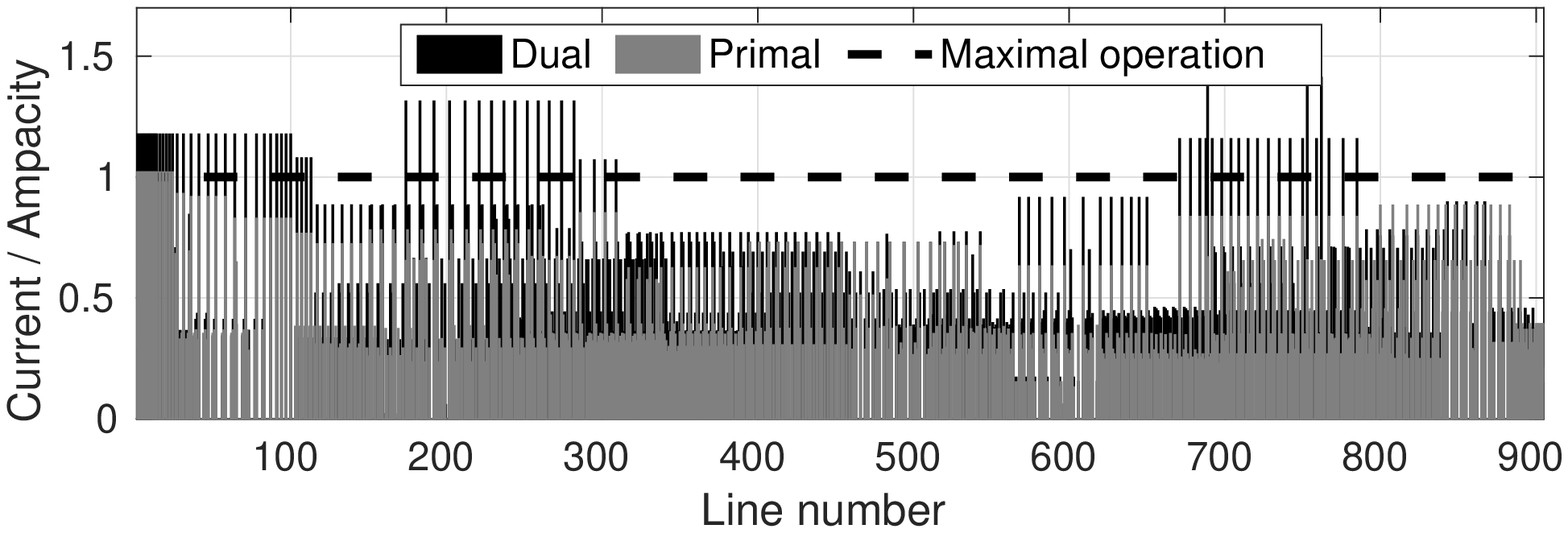}
  \caption{EV NUM control (voltages and currents at peak time).}
  \label{fig:control}
\end{subfigure}
\caption{EV charging effects on the IEEE European Low Voltage Test Feeder.}
\label{fig:dynamic}
\end{figure*}

To evaluate the effects of EV charging on the IEEE European Low Voltage Test Feeder, we conduct simulations using GridLab-D~\cite{chassin2008gridlab}. 
First, we evaluate the effect of charging the EVs without control for different charging rates. 
The results in Fig. \ref{fig:no_control} reveal that our evaluation grid can support the charging of EVs with a maximum charging power of 4 kW without any charging control. 
However, for a charging power of 7 kW, we start to see ampacity and voltage violations. 
With a charging power of 20 kW, we also see violations of the transformer's maximal loading. 
Hence, real-time EV charging control is required to allow higher charging powers than 4 kW and make better use of the grid's capacity without violating its constraints.
This experiment also revealed that for this particular feeder ampacity violations happen before voltage violations as the charging power of the EVs increases. 

We evaluated the impact of the EV NUM dual and primal algorithms on the grid by running simulations using the load results of our dynamic evaluation in Section \ref{sec:dynamic_evaluation}. 
Fig. \ref{fig:control} demonstrates that the dual algorithm violates the network constraints, whereas the primal algorithm is able to guarantee charging rates that remain within the network constraints. 
Our GridLab-D simulation results match the behavior obtained in Section \ref{sec:dynamic_evaluation} with the EV NUM formulation. 
Hence, our experiments show that the EV NUM problem can capture the relevant constraints to design an effective real-time EV control algorithm for distribution grids, where the ampacity violations are the limiting factor. 
Moreover, this result clearly demonstrates the effectiveness of the primal algorithm for real-time EV charging control.



\section{Discussion} 
\label{sec:discussion}


\subsection{EV NUM  for real-time EV charging control}

The EV NUM problem formulation offers a simplified model for the real-time EV charging problem. 
The goal of using EV NUM is to capture the simplest form of the real-time EV charging problem that still encompasses the relevant features required to design a distributed control algorithm. 
Several arguments can be made for the use of a more comprehensive model. However, more complex formulations do not allow the formulation of efficient distributed algorithms, that can cope with the real-time requirements. Another argument in favor of the simple EV NUM model is that more complex models need more network data. When it comes to massive application of the control algorithm, it is not certain that accurate network data of the distribution grids is available. In fact, the accuracy and even the availability of accurate distribution network models is known to be an issue \cite{EsriBook}. 
Therefore, we consider that the EV NUM problem formulation can be an alternative to more complex models with higher computational demands and that require highly accurate network data.
%


\subsection{Dual vs primal EV NUM}

Our evaluations have shown that the primal algorithm has an advantage over the dual algorithm regarding both scalability and reliability. 
The scalability advantage results from the stability issues of the dual algorithm. 
Our static evaluations in Section \ref{sec:static_evaluation} confirm that the dual algorithm's stability depends on the step size. 
As explained in Section \ref{sec:convergence_dual}, the theoretical maximum stable step size is inversely proportional to the number of EVs and the size of the distribution grid. 
Hence, as the size of the problem increases, the step size must be reduced to guarantee stability, which in turn increases the number of iterations required to reach convergence. 
The primal algorithm does not need to reduce its step size as the problem size increases and therefore scales better to larger problems. 
Regarding the reliability advantage, our dynamic evaluations of Section \ref{sec:dynamic_evaluation} show that the dual algorithm requires a higher update frequency than the primal algorithm to avoid the violation of grid constraints. 
The reliability advantage of the primal algorithm comes from its anytime property, which guarantees that the problem constraints are satisfied on each iteration.

Nevertheless, the dual algorithm might be the preferred option when the problem size is relatively small and minor violations of the grid constraints are permitted. 
The experiments in Section \ref{sec:evaluation_effect} clearly show that the dual algorithm allows the EVs to charge faster at the cost of grid constraint violations. 


\section{Conclusions} 
\label{conclusion}

%


This paper proposes the use of the NUM formulation to formally design real-time EV charging control protocols, compares the resulting dual and primal distributed solution algorithms and provides a comprehensive evaluation of their impact when implemented on the IEEE European Low Voltage Test Feeder. 
Our experimental evaluation demonstrates the effectiveness of the NUM formulation to model networks where the ampacity constraints are the main bottleneck. Our results also show that the primal algorithm outperforms the dual algorithm regarding scalability and reliability. 

\bibliographystyle{IEEEtran}
\bibliography{litCDC2014_extension}

\begin{thebibliography}{10}
\providecommand{\url}[1]{#1}
\csname url@samestyle\endcsname
\providecommand{\newblock}{\relax}
\providecommand{\bibinfo}[2]{#2}
\providecommand{\BIBentrySTDinterwordspacing}{\spaceskip=0pt\relax}
\providecommand{\BIBentryALTinterwordstretchfactor}{4}
\providecommand{\BIBentryALTinterwordspacing}{\spaceskip=\fontdimen2\font plus
\BIBentryALTinterwordstretchfactor\fontdimen3\font minus
  \fontdimen4\font\relax}
\providecommand{\BIBforeignlanguage}[2]{{%
\expandafter\ifx\csname l@#1\endcsname\relax
\typeout{** WARNING: IEEEtran.bst: No hyphenation pattern has been}%
\typeout{** loaded for the language `#1'. Using the pattern for}%
\typeout{** the default language instead.}%
\else
\language=\csname l@#1\endcsname
\fi
#2}}
\providecommand{\BIBdecl}{\relax}
\BIBdecl

\bibitem{EVintegration}
M.~Bradley and Associates, ``{Electric vehicle grid integration in the U.S.,
  Europe, and China},'' M.J Bradley and Associates, Tech. Rep., 2013.

\bibitem{EVChargingProblem}
\BIBentryALTinterwordspacing
C.~Sullivan. (2009, August) {Will Electric Cars Wreck the Grid?} Scientific
  American. [Online]. Available:
  \url{http://scientificamerican.com/article/will-electric-cars-wreck-the-grid}
\BIBentrySTDinterwordspacing

\bibitem{putrus2009impact}
G.~Putrus, P.~Suwanapingkarl, D.~Johnston, E.~Bentley, and M.~Narayana,
  ``{Impact of electric vehicles on power distribution networks},'' in
  \emph{{Vehicle Power and Propulsion Conference, 2009. VPPC'09. IEEE}}.\hskip
  1em plus 0.5em minus 0.4em\relax IEEE, 2009, pp. 827--831.

\bibitem{richardson2010impact}
P.~Richardson, D.~Flynn, and A.~Keane, ``{Impact assessment of varying
  penetrations of electric vehicles on low voltage distribution systems},'' in
  \emph{{Power and Energy Society General Meeting, 2010 IEEE}}.

\bibitem{gomez2003impact}
J.~C. Gomez and M.~M. Morcos, ``{Impact of EV battery chargers on the power
  quality of distribution systems},'' \emph{Power Delivery, IEEE Transactions
  on}, vol.~18, no.~3, pp. 975--981, 2003.

\bibitem{schroeder2012economics}
A.~Schroeder and T.~Traber, ``{The economics of fast charging infrastructure
  for electric vehicles},'' \emph{Energy Policy}, vol.~43, pp. 136--144, 2012.

\bibitem{botsford2009fast}
C.~Botsford and A.~Szczepanek, ``{Fast charging vs. slow charging: Pros and
  cons for the new age of electric vehicles},'' in \emph{{International Battery
  Hybrid Fuel Cell Electric Vehicle Symposium}}, 2009.

\bibitem{willis2004power}
H.~L. Willis, \emph{{Power distribution planning reference book}}.\hskip 1em
  plus 0.5em minus 0.4em\relax CRC press, 2004.

\bibitem{EnergyInformatics}
C.~Goebel, H.-A. Jacobsen, V.~Razo, C.~Doblander, J.~Rivera \emph{et~al.},
  ``\BIBforeignlanguage{English}{{Energy Informatics}},''
  \emph{\BIBforeignlanguage{English}{Business \& Information Systems
  Engineering}}, 2013.

\bibitem{ardakanian2012realtime}
O.~Ardakanian, C.~Rosenberg, and S.~Keshav, ``{Real-time distributed congestion
  control for electrical vehicle charging},'' \emph{ACM SIGMETRICS Performance
  Evaluation Review}, vol.~40, no.~3, pp. 38--42, 2012.

\bibitem{Ardakanian2013a}
------, ``{Distributed control of electric vehicle charging},'' in
  \emph{{Proceedings of the fourth International Conference on Future Energy
  Systems}}, 2013.

\bibitem{ardakanian2014real}
O.~Ardakanian, S.~Keshav, and C.~Rosenberg, ``{Real-Time Distributed Control
  for Smart Electric Vehicle Chargers: From a Static to a Dynamic Study},''
  \emph{IEEE Transactions on Smart Grid}, vol.~5, no.~5, pp. 2295--2305, Sept
  2014.

\bibitem{JoseRiveraCDC14}
J.~Rivera and H.-A. Jacobsen, ``{A distributed anytime algorithm for network
  utility maximization with application to real-time EV charging control},'' in
  \emph{{Decision and Control (CDC), 2014 IEEE 53rd Annual Conference on}}, Dec
  2014, pp. 947--952.

\bibitem{RiveraEenergy}
J.~Rivera, C.~Goebel, and H.-A. Jacobsen, ``{A distributed anytime algorithm
  for real-time EV charging congestion control},'' in \emph{{Proceedings of the
  2015 ACM Sixth International Conference on Future Energy Systems}}, ser.
  {e-Energy '15}.

\bibitem{turitsyn2010robust}
K.~Turitsyn, N.~Sinitsyn, S.~Backhaus, and M.~Chertkov, ``{Robust
  broadcast-communication control of electric vehicle charging},'' in
  \emph{{Smart Grid Communications (SmartGridComm), 2010 First IEEE
  International Conference on}}.\hskip 1em plus 0.5em minus 0.4em\relax IEEE,
  2010, pp. 203--207.

\bibitem{studli2012aimd}
``{AIMD-like algorithms for charging electric and plug-in hybrid vehicles},''
  in \emph{{Electric Vehicle Conference (IEVC), 2012 IEEE International}}.

\bibitem{Gan.2012}
L.~Gan, U.~Topcu, and S.~H. Low, ``{Optimal decentralized protocol for electric
  vehicle charging},'' \emph{IEEE Trans. Power Systems}, vol.~PP, no.~99, pp.
  1--12, 2012.

\bibitem{Mercurio2013}
A.~Mercurio, A.~{Di Giorgio}, and F.~Purificato, ``{Optimal fully electric
  vehicle load balancing with an ADMM algorithm in smart grids},'' in
  \emph{{Control \& Automation (MED), 2013 21st Mediterranean Conference
  on}}.\hskip 1em plus 0.5em minus 0.4em\relax IEEE, 2013, pp. 119--124.

\bibitem{Ma2014}
W.-J. Ma, V.~Gupta, and U.~Topcu, ``{On distributed charging control of
  electric vehicles with power network capacity constraints},'' in
  \emph{{American Control Conference (ACC), 2014}}.\hskip 1em plus 0.5em minus
  0.4em\relax IEEE, 2014, pp. 4306--4311.

\bibitem{JoseRivera}
J.~Rivera, P.~Wolfrum, S.~Hirche, C.~C. Goebel, and H.-A. Jacobsen,
  ``{Alternating direction method of multipliers for decentralized electric
  vehicle charging control},'' in \emph{{52nd IEEE Conference on Decision and
  Control (CDC)}}, 2013, p.~6.

\bibitem{RiveraEVADMM}
J.~Rivera, C.~Goebel, and H.~A. Jacobsen, ``{Distributed convex optimization
  for electric vehicle aggregators},'' \emph{IEEE Transactions on Smart Grid},
  vol.~PP, no.~99, pp. 1--12, 2016.

\bibitem{Kelly98}
F.~Kelly, A.~Maulloo, and D.~Tan, ``{Rate control in communication networks:
  Shadow prices, proportional fairness and stability},'' in \emph{{Journal of
  the Operational Research Society}}, vol.~49, 1998.

\bibitem{mo2000fair}
J.~Mo and J.~Walrand, ``{Fair end-to-end window-based congestion control},''
  \emph{IEEE/ACM Transactions on Networking (ToN)}, vol.~8, no.~5, pp.
  556--567, 2000.

\bibitem{BertsekasDCBook}
D.~P. Bertsekas and J.~N. Tsitsiklis, \emph{{Parallel and Distributed
  Computation: Numerical Methods}}.\hskip 1em plus 0.5em minus 0.4em\relax
  Upper Saddle River, NJ, USA: Prentice-Hall, Inc., 1989.

\bibitem{LectureBoyd}
S.~Boyd, L.~Xiao, A.~Mutapic, J.~Dattorro, and J.~Mattingley, ``{Subgradient
  methods, decomposition methods, alternating projections},'' {Notes for
  EE364b, Stanford University}, 2007.

\bibitem{BoydBook}
S.~P. Boyd and L.~Vandenberghe, \emph{{Convex Optimization}}.\hskip 1em plus
  0.5em minus 0.4em\relax Cambridge University Press, 2004.

\bibitem{EULVgrid}
{IEEE}, ``{IEEE Radial Distribution Test Feeders},'' Online:
  \href{https://ewh.ieee.org/soc/pes/dsacom/testfeeders/}{https://ewh.ieee.org/soc/pes/dsacom/testfeeders/},
  accessed: 15.02.2016.

\bibitem{IEEERadialTestFeeders}
W.~Kersting, ``{Radial distribution test feeders},'' in \emph{{Power
  Engineering Society Winter Meeting}}, vol.~2, 2001.

\bibitem{chassin2008gridlab}
D.~P. Chassin, K.~Schneider, and C.~Gerkensmeyer, ``{GridLAB-D: An open-source
  power systems modeling and simulation environment},'' in \emph{{2008 IEEE/PES
  Transmission and Distribution Conference and Exposition}}, 2008.

\bibitem{EsriBook}
B.~Meehan, \emph{{Modeling Electric Distribution with GIS}}.\hskip 1em plus
  0.5em minus 0.4em\relax Esri Press, 2013.

\end{thebibliography}

%


%
%
%
%
%




\end{document}